\documentclass[12pt]{iopart}

\begin{document}

\title{Soliton Solutions of Fractional order KdV-Burger's Equation}

\author{Muhammad Younis$^{1,*}$}

\address{$^{1}$Center for Undergraduate Studies, University of the Punjab, 54590 Lahore, Pakistan.}
\ead{$^{*}$younis.pu@gmail.com}
\begin{abstract}
In this article, the new exact travelling wave solutions of the time- and space-fractional KdV-Burgers equation has been found. For this the fractional complex transformation have been implemented to convert nonlinear partial fractional differential equations to nonlinear ordinary differential equations, in the sense of the Jumarie's modified Riemann-Liouville derivative. Afterwards, the improved $(G^{'}/G)$-expansion method can be implemented to celebrate the soliton solutions of KdV-Burger's equation of fractional order.

\end{abstract}


\section{Introduction}
Nonlinear differential equations involving derivatives
of fractional order have shown to be adequate models for
many important phenomena in electromagnetic, acoustics,
electrochemistry, cosmology, biological and material science \cite{WG95, PI99}.
Fractional differential equations can be considered as the generalization
form of the classical differential equations.

In the literature, the different techniques exist to find the
soliton and travelling wave solutions of nonlinear evolution
equations. For example, the $(G^{'}/G)$-expansion method
\cite{MW08}, the first integral method \cite{MY13}, method of
variation of parameters \cite{SS11}. Jacobi elliptic function
expansion method \cite{SK01}, tanh-function method \cite{EJ96},
homotopy perturbation method \cite{KG11} \textit{etc.} can also be
used for solving the nonlinear evolution equations.

In this article, the $(G^{'}/G)$-expansion method \cite{MW08} has been applied to find the new exact traveling wave solutions of the nonlinear time and space fractional order Korteweg-de Vries-Burger's equation (KdV-Burger's equation), given in equation (1), in the sense of the Jumarie's modified Riemann-Liouville derivative \cite{GJ06} of order $\alpha$, defined by the following expression
\begin{eqnarray*}
D_{s}^{\alpha}f(s)= \left\{\begin{array}{rr}
\frac{1}{\Gamma(1-\alpha)}\frac{d}{ds}\int_{0}^{s}(s-\xi)^{-\alpha}(f(\xi)-f(0))d\xi,~~~~~0<\alpha<1,\\
(f^{(n)}(s))^{\alpha-n},~~~~~~~~~~~~~~~~~~~~~~~ n\leq \alpha < n+1, n\geq 1.
\end{array}\right.
\end{eqnarray*}
Moreover, some properties for the modified Riemann-Liouville derivative have also been given as follows
\begin{eqnarray*}
&&D_{s}^{\alpha}s^r=\frac{\Gamma(1+r)}{\Gamma(1+r-\alpha)}s^{r-\alpha},\\
&&D_{s}^{\alpha}(f(s)g(s))=f(s)D_{s}^{\alpha}g(s)+g(s)D_{s}^{\alpha}t(s),\\
&&D_{s}^{\alpha}f[g(s)]=f^{'}_{g}[g(s)]D_{s}^{\alpha}g(s)=D_{s}^{\alpha}f[g(s)](g^{'}(t))^{\alpha}.
\end{eqnarray*}
The nonlinear space-time fractional Korteweg-de Vries-Burgers (KdV-Burgers) equation \cite{QW06,ZF02}, can be given as follows
\begin{eqnarray}
\frac{\partial^{\alpha}u}{\partial t^{\alpha}}+\omega u\frac{\partial^{\beta}u}{\partial x^{\beta}}+\eta \frac{\partial^{2\beta}u}{\partial x^{2\beta}}+\nu \frac{\partial^{3\beta}u}{\partial x^{3\beta}}=0,~~t>0,~0<\alpha, ~\beta \leq 1.
\end{eqnarray}
It is applied as a nonlinear model of the propagation of waves on an elastic tube filled with a viscous fluid \cite{RS70}.
The equation (1) can be treated as the space-time fractional Korteweg-de Vries (KdV) equation, with the choice $\omega \neq 0, \nu \neq 0$ and $\eta = 0$.
\begin{eqnarray}
\frac{\partial^{\alpha}u}{\partial t^{\alpha}}+\omega u\frac{\partial^{\beta}u}{\partial x^{\beta}}+\nu \frac{\partial^{3\beta}u}{\partial x^{3\beta}}=0,~~t>0,~0<\alpha, ~\beta \leq 1.
\end{eqnarray}
The exact solution for the equation (2) has been found \cite{KG12}.

The rest of the letter is organized as follows. In section 2, the $(G^{'}/G)$-expansion method has been proposed to find the exact solutions of nonlinear evolution equation with the help of fractional complex transformation. As an application, the new exact travelling wave solutions of KdV-Burger's equation has been found in section 3. In last section 4, the conclusion has been drawn.
\section{The $(G^{'}/G)$-expansion method for  nonlinear fractional partial differential equations}
~~~~In this section, the {$(G^{'}/G)$-expansion method has been discussed to obtain the solutions of nonlinear fractional order partial differential equations (NFPDEs).\\
For this, we consider the following NFPDE
\begin{eqnarray}
&&P\left(u,D_{t}^{\alpha}u,D_{s}^{\beta}u,D_{x}^{\gamma}u,...,D_{t}^{\alpha}D_{t}^{\alpha}u,D_{t}^{\alpha}D_{s}^{\beta}u                   ,D_{s}^{\beta}D_{s}^{\beta}u,D_{s}^{\beta}D_{x}^{\gamma}u,...\right)=0,~~~~~~~~~~~~~~\\
&&~~~~~~~~~~~~~~~~~~~~~~~~~~~~~~~~~~~~~~~~~~~~~~~~~~~~~~~~~~~~~~for~~0<\alpha, \beta, \gamma <1,\nonumber
\end{eqnarray}
where $u$ is an unknown function and $P$ is a polynomial of $u$ and its partial fractional derivatives along with the involvement of higher order derivatives and nonlinear terms.\\
To find the exact solutions, the following steps can be performed.\\
\textbf{Step 1} First, we convert the NFPDE into nonlinear ordinary differential equations using fractional complex transformation introduced by Li \textit{et al.} \cite{ZL10}.
The travelling wave variable
\begin{eqnarray}
u(t,x,y)=u(\xi),~~
\xi=\frac{Kt^{\alpha}}{\Gamma(\alpha+1)}+\frac{Lx^{\beta}}{\Gamma(\beta+1)}+\frac{My^{\gamma}}{\Gamma(\gamma+1)}
\end{eqnarray}
where $K, L$ and $M$ are non-zero arbitrary constants, permits to reduce equation (3) to an ODE of $u=u(\xi)$ in the following form
\begin{eqnarray}
P(u,u^{'},u^{''},u^{'''},...)=0.
\end{eqnarray}
If the possibility occurs, the above equation can be integrated term by term once or more times.\\
\textbf{Step 2} Suppose that the solution of equation (5) can be expressed as a polynomial of {$(G^{'}/G)$ in the following form
\begin{eqnarray}
u(\xi)=\sum_{i=-m}^{m}\alpha _{i}\left(\frac{G^{'}}{G}\right)^{i}, ~~~\alpha_{m}\neq0,
\end{eqnarray}
where $\alpha_i'$s are constants and $G(\xi)$ satisfies the following second order linear ordinary differential equation.
\begin{eqnarray}
G^{''}(\xi)+\lambda G^{'}(\xi)+\mu G(\xi)=0,
\end{eqnarray}
with $\lambda$ and $\mu$ as constants.\\
\textbf{Step 3} The homogeneous balance can be used, to determine the positive integer $m$, between the highest order derivatives and the nonlinear terms appearing in (5).\\
Moreover, the degree of $u(\xi)$ can be defined as $D[u(\xi)]=m$, which gives rise to the degree of the other expressions as follows
\begin{eqnarray*}
&&D\left[\frac{d^{q}u}{d\xi q}\right]=m+q,\\
&&D\left[u^p\left(\frac{d^{q}u}{d\xi q}\right)^s\right]=mq+s(q+m).
\end{eqnarray*}
Therefore, the value of $m$ can be obtained for the equation (6).\\
\textbf{Step 4} After the substitution of equation (6) into equation (5) and using equation (7), we collect all the terms with the same order of
{$(G^{'}/G)$ together. 
Equate each coefficient of the obtained polynomial to zero, yields the set of algebraic equations for $K, L, M, \lambda, \mu$ and $\alpha_i (i=0, \pm1, \pm2,...,\pm m)$.\\
\textbf{Step 5} After solving the system of algebraic equations, and using equation (3), the variety of exact solutions can be constructed.
\section{Application of Fractional KdV-Burgers equation}
In this section, the improved {$(G^{'}/G)$-expansion method have been used to construct the exact solutions for  nonlinear space-time fractional KdV-Burgers equation (1).
\begin{eqnarray}
\frac{\partial^{\alpha}u}{\partial t^{\alpha}}+\omega u\frac{\partial^{\beta}u}{\partial x^{\beta}}+\eta \frac{\partial^{2\beta}u}{\partial x^{2\beta}}+\nu \frac{\partial^{3\beta}u}{\partial x^{3\beta}}=0,~~t>0,~0<\alpha, ~\beta \leq 1.
\end{eqnarray}
It can be observed that the fractional complex transform
\begin{eqnarray}
u(x,t)=u(\xi),~~
\xi=\frac{Kx^{\beta}}{\Gamma(\beta+1)}+\frac{Lt^{\alpha}}{\Gamma(\alpha+1)}
\end{eqnarray}
where $K$ and $L$
are constants, permits to reduce the equation (8) into an ODE. After integrating once, we have the following form
\begin{eqnarray}
C+LU+\frac{1}{2}\omega KU^2+\eta K^2 U^{'}+\nu K^3 U^{''}=0,
\end{eqnarray}
where $C$ is a constant of integration. Now by considering the homogeneous balance between the highest order derivatives and nonlinear term presented in equation (10), we have the following form
\begin{eqnarray}
u(\xi)=\sum_{i=-2}^{2}\alpha _{i}\left(\frac{G^{'}}{G}\right)^{i}, ~~~\alpha_{2}\neq0,
\end{eqnarray}
where $\alpha_{-2}, \alpha_{-1}, \alpha_{0}, \alpha_{1}, \alpha_{2}, K$ and $L$ are arbitrary constants. To determine these constants substitute the equation(11) into (10), and collect all the terms with the same power of $G^{'}/G$ together. After equating each coefficient equal to zero, yields a set of algebraic equations. After solving these algebraic equations with the help of software \textit{Maple}, the following results can be yield.\\
\textbf{Case 1} For the values
\begin{eqnarray*}
C=\frac{(\lambda \eta K^2-L^2)}{2K}-\frac{\eta \lambda K}{\omega}(\lambda \eta K^2-L),~~\alpha_0=\frac{\lambda \eta K^2-L}{K\omega},~~
\alpha_1=\frac{2\eta K}{\omega}\\~~and ~~~\nu=\alpha_{-2}=\alpha_{-1}=\alpha_{2}=0.
\end{eqnarray*}
It can, thus, be written equation (11) as
\begin{eqnarray}
u(\xi)=\frac{2\eta K}{\omega}\left(\frac{G^{'}}{G}\right)+\frac{\lambda \eta K^2-L}{K\omega}.
\end{eqnarray}
From equations (7) and (12), the following travelling wave solutions can be obtained as\\
If $\lambda^2-4\mu > 0$, then we have the following hyperbolic solution
\begin{eqnarray}
u_{1}(\xi)=\frac{2\eta K}{\omega}\sqrt{\lambda^2-4\mu}\left(\frac{A \cosh(\frac{\xi}{2}\sqrt{\lambda^2-4\mu})+B \sinh(\frac{\xi}{2}\sqrt{\lambda^2-4\mu})}{A \sinh(\frac{\xi}{2}\sqrt{\lambda^2-4\mu})+B \cosh(\frac{\xi}{2}\sqrt{\lambda^2-4\mu})}\right)\nonumber\\
+\frac{\lambda \eta K^2-L}{K\omega}.
\end{eqnarray}
If $\lambda^2-4\mu < 0$, then we have the following trigonometric solution
\begin{eqnarray}
u_{2}(\xi)=&&\frac{2\eta K}{\omega}\sqrt{4\mu-\lambda^2}\left(\frac{-A \sin(\frac{\xi}{2}\sqrt{4\mu-\lambda^2})+B \cos(\frac{\xi}{2}\sqrt{4\mu-\lambda^2})}{A \cos(\frac{\xi}{2}\sqrt{4\mu-\lambda^2})+B \sin(\frac{\xi}{2}\sqrt{4\mu-\lambda^2})}\right)\nonumber\\
&&+\frac{\lambda \eta K^2-L}{K\omega}
\end{eqnarray}
and if $\lambda^2-4\mu =0$, then we have the following solution
\begin{eqnarray}
u_{3}(\xi)=\frac{2\eta K}{\omega}\left(\frac{B \xi}{A +B \xi}\right)+\frac{\lambda \eta K^2-L}{K\omega},
\end{eqnarray}
~~~~~~~~~~~~where
\begin{eqnarray*}
\xi=\frac{Kx^{\beta}}{\Gamma(\beta+1)}+\frac{Lt^{\alpha}}{\Gamma(\alpha+1)}.
\end{eqnarray*}
In particular, if we take $A=0$ in equation (13), then it takes the form
\begin{eqnarray}
u_{4}(\xi)=\frac{2\eta K}{\omega}\sqrt{\lambda^2-4\mu}\left( \tanh(\frac{\xi}{2}\sqrt{\lambda^2-4\mu})\right)+\frac{\lambda \eta K^2-L}{K\omega}.
\end{eqnarray}
In particular, if we take $B=0$ in equation (14), then it takes the form
\begin{eqnarray}
u_{5}(\xi)=-\frac{2\eta K}{\omega}\sqrt{4\mu-\lambda^2}\left( \tan(\frac{\xi}{2}\sqrt{4\mu-\lambda^2})\right)+\frac{\lambda \eta K^2-L}{K\omega}.
\end{eqnarray}
\\
\textbf{Case 2} For the values
\begin{eqnarray*}
C=\frac{(\alpha_0^2+3\alpha_1^2\mu-\mu^2\alpha_1^2)\eta K^2}{\alpha_1},~~\alpha_1=\frac{2\eta K}{\omega},~~\alpha_{-1}=-\frac{2\eta \mu K}{\omega}\\
\alpha_0=\frac{-L}{K\omega}~~and~~\nu =\lambda =\alpha_{-2}=\alpha_{2}=0.
\end{eqnarray*}
It can, thus, be written equation (11) as
\begin{eqnarray}
u(\xi)=\frac{2\eta K}{\omega}\left(\frac{G^{'}}{G}\right)-\frac{L}{K\omega}-\frac{2\eta \mu K}{\omega}\left(\frac{G^{'}}{G}\right)^{-1},
\end{eqnarray}
From the equations (7) and (18), we have the following travelling wave solutions.\\
For $\lambda^2-4\mu > 0$, the following hyperbolic solution can be obtained
\begin{eqnarray}
u_6(\xi)=\frac{2\eta K}{\omega}\sqrt{\lambda^2-4\mu}\left(\frac{A \cosh(\frac{\xi}{2}\sqrt{\lambda^2-4\mu})+B \sinh(\frac{\xi}{2}\sqrt{\lambda^2-4\mu})}{A \sinh(\frac{\xi}{2}\sqrt{\lambda^2-4\mu})+B \cosh(\frac{\xi}{2}\sqrt{\lambda^2-4\mu})}\right)\nonumber\\-\frac{L}{K\omega}-\frac{2\eta \mu K}{\omega}(\lambda^2-4\mu)^{-\frac{1}{2}}\left(\frac{A \cosh(\frac{\xi}{2}\sqrt{\lambda^2-4\mu})+B \sinh(\frac{\xi}{2}\sqrt{\lambda^2-4\mu})}{A \sinh(\frac{\xi}{2}\sqrt{\lambda^2-4\mu})+B \cosh(\frac{\xi}{2}\sqrt{\lambda^2-4\mu})}\right)^{-1}.\nonumber\\
~~~~
\end{eqnarray}
If $\lambda^2-4\mu < 0$, then we have the following trigonometric solution
\begin{eqnarray}
u_{7}(\xi)=\frac{2\eta K}{\omega}\sqrt{4\mu-\lambda^2}\left(\frac{-A \sin(\frac{\xi}{2}\sqrt{4\mu-\lambda^2})+B \cos(\frac{\xi}{2}\sqrt{4\mu-\lambda^2})}{A \cos(\frac{\xi}{2}\sqrt{4\mu-\lambda^2})+B \sin(\frac{\xi}{2}\sqrt{4\mu-\lambda^2})}\right)\nonumber\\-\frac{L}{K\omega}-\frac{2\eta \mu K}{\omega}(4\mu-\lambda^2)^{-\frac{1}{2}}\left(\frac{-A \sin(\frac{\xi}{2}\sqrt{4\mu-\lambda^2})+B \cos(\frac{\xi}{2}\sqrt{4\mu-\lambda^2})}{A \cos(\frac{\xi}{2}\sqrt{4\mu-\lambda^2})+B \sin(\frac{\xi}{2}\sqrt{4\mu-\lambda^2})}\right)^{-1}\nonumber\\
~~~~
\end{eqnarray}
and if $\lambda^2-4\mu =0$, then we have the following solution
\begin{eqnarray}
u_8(\xi)=\frac{2\eta K}{\omega}\left(\frac{B \xi}{A +B \xi}\right)-\frac{L}{K\omega}-\frac{2\eta \mu K}{\omega}\left(\frac{B \xi}{A +B \xi}\right)^{-1},
\end{eqnarray}
~~~~~~~~~~~~where
\begin{eqnarray*}
\xi=\frac{Kx^{\beta}}{\Gamma(\beta+1)}+\frac{Lt^{\alpha}}{\Gamma(\alpha+1)}.
\end{eqnarray*}
In particular, if we take $A=0$ in equation (19), then the following solution can be obtained
\begin{eqnarray}
u_9(\xi)=&&\frac{2\eta K}{\omega}\sqrt{\lambda^2-4\mu}\left(\tanh(\frac{\xi}{2}\sqrt{\lambda^2-4\mu})\right)-\frac{L}{K\omega}
\nonumber\\&&-\frac{2\eta \mu K}{\omega}(\lambda^2-4\mu)^{-\frac{1}{2}}\left( \tanh(\frac{\xi}{2}\sqrt{\lambda^2-4\mu})\right)^{-1}.
\end{eqnarray}
In particular, if we take $B=0$ in equation (20), then the following solution can be obtained
\begin{eqnarray}
u_{10}(\xi)=&&\frac{2\eta \mu K}{\omega}(4\mu-\lambda^2)^{-\frac{1}{2}}\left(\tan(\frac{\xi}{2}\sqrt{4\mu-\lambda^2})\right)^{-1}\nonumber\\
&&-\frac{2\eta K}{\omega}\sqrt{4\mu-\lambda^2}\left(\tan(\frac{\xi}{2}\sqrt{4\mu-\lambda^2})\right)-\frac{L}{K\omega}.
\end{eqnarray}
These are the required soliton solutions.

\section{Conclusion}
The improved $(G^{'}/G)$-expansion method has been extended to solve the nonlinear fractional partial differential equation using the fractional complex transformation.
As application, the new exact travelling wave solutions for the space-time fractional KdV-Burger's equation have been found.
It can be concluded that this method is very simple, reliable and propose a variety of exact solutions to evolution equations of fractional order.

\section*{References}

%
%
%
%

\end{document}